\documentclass[a4paper]{spie}  %>>> use this instead for A4 paper
\usepackage[]{graphicx}
\usepackage{amssymb,amsmath,psfrag,url}

\title{Numerical Investigation of Photonic Crystal Microcavities in Silicon-on-Insulator Waveguides}

\author{
Sven Burger,\supit{\,ab}
Frank Schmidt\supit{\,ab}
Lin Zschiedrich,\supit{\,ab}
\skiplinehalf
\supit{a}
Zuse Institute Berlin (ZIB),
Takustra{\ss}e 7,
D\,--\,14\,195 Berlin,
Germany
\smallskip\\
\supit{b}
JCMwave GmbH,
Bolivarallee 22, 
D\,--\,14\,050 Berlin,
Germany
}
\authorinfo{
Corresponding author: S. Burger\\
URL: http://www.zib.de/Numerik/NanoOptics/\\
URL: http://www.jcmwave.com\\
Email: burger@zib.de
}

\begin{document}
\maketitle
%\today

%%%%%%%%%%%%%%%%%%%%%%%%%%%%%%%%%%%%%%%%%%%%%%%%%%%%%%%%%%%%% 
%% SPIE Copyright form 

%Proc. SPIE, Vol. 7609, 76091Q (2010)
\noindent
This paper will be published in Proc.~SPIE Vol. {\bf 7609}
(2010) 76091Q,  
({\it Photonic and Phononic Crystal Materials and Devices X, Ali Adibi, Shawn-Yu Lin, Axel Scherer, Editors})
and is made available as an electronic preprint with permission of SPIE. 
Copyright 2010 Society of Photo-Optical Instrumentation Engineers. 
One print or electronic copy may be made for personal use only. 
Systematic reproduction and distribution, duplication of any material in this paper for a fee or for 
commercial purposes, or modification of the content of the paper are prohibited. 

%%%%%%%%%%%%%%%%%%%%%%%%%%%%%%%%%%%%%%%%%%%%%%%%%%%%%%%%%%%%% 

\begin{abstract}
Optical resonances in 1D photonic crystal microcavities are investigated numerically using
finite-element light scattering and eigenmode solvers. 
The results are validated by comparison to experimental and theoretical findings from 
the literature. 
The influence of nanometer-scale geometry variations 
on the resonator performance is studied. 
Limiting factors to ultra-high Q-factor performance are identified. 
\end{abstract}

\keywords{optical microcavity, nanooptics, integrated optics, 3D Maxwell solver, finite-element method}

\section{Introduction}

\begin{figure}[b]
\begin{center}
  \includegraphics[width=.8\textwidth]{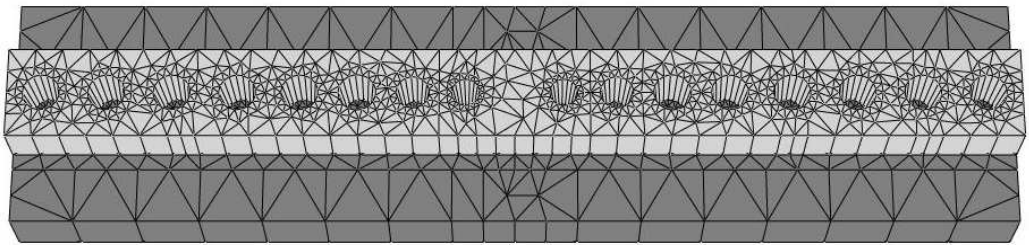}
  \caption{Geometry of a PhC microcavity, discretized with a prismatoidal mesh. 
Dark grey: Silica substrate, 
light grey: Silicon waveguide, 
perforated by air cones (side-wall angle of 85 degrees). 
The structure is surrounded by air, substrate and waveguides extend to infinity. 
}
\label{fig_grid}
\end{center}
\end{figure}

Integration of optical components with electronic circuits promises 
to provide increased functionality. 
Building blocks of such integrated optical devices are wave\-guides, 
modulators, switches, and others. 
Due to their high quality factor (Q-factor) and small modal volume,  
one dimensional photonic crystal (PhC) microcavities 
in silicon-on-insulator 
are candidates for integrated all-optical switching devices. 
High Q-factors have been reported for such devices by Zain {\it et al}~\cite{Zain2008a}. 
Further realized designs of optical microcavities include 
2D photonic crystal resonators~\cite{Tanaka2008a}, suspended air-bridge cavities~\cite{Ripin2000a}
and circular grating resonators~\cite{Schoenenberger2009a}.

In this contribution we numerically investigate high-Q resonances in PhC microcavities, where we 
partially review results of an earlier contribution~\cite{Burger2009Tacona}. 
We analyse the photonic devices through numerical
simulations of light propagation using a time-harmonic finite-element method 
(FEM) light scattering solver. 
The method is validated by comparison to experimental and theoretical 
results, and the effect of small geometry variations on device performance 
is investigated. 
We further use an eigenmode solver for direct computation of resonance wavelengths and 
Q-factors. 
The light scattering solver, the eigenmode solver as well as a propagation mode solver 
used for computing incident waveguide modes are parts of the commercial FEM programme package 
JCMsuite, which is developed by JCMwave and ZIB~\cite{Burger2008ipnra}.
The solvers enable fast and accurate simulations of main properties of optical 
microcavities.

\begin{figure}[t]
\begin{center}
  \includegraphics[width=.8\textwidth]{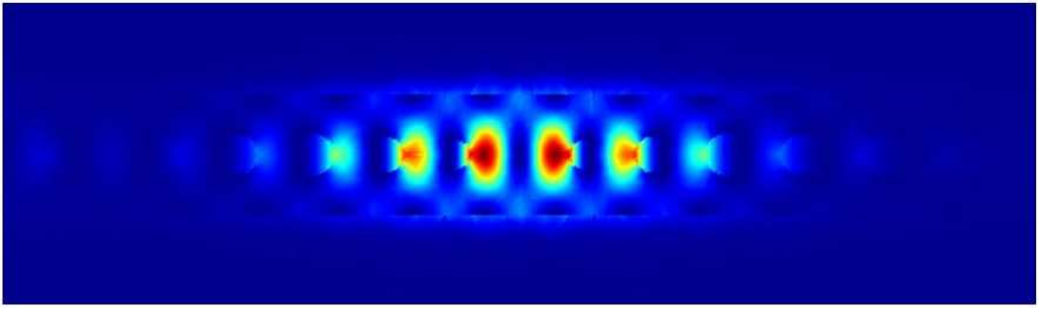}
  \caption{
Electric field intensity distribution in the microcavity close to a  resonance.
Cross-section at $z=h/2$ above the substrate through the 3D solution in 
pseudocolor representation (blue/dark: low intensity,  red/bright: high intensity).
}
\label{fig_solution}
\end{center}
\end{figure}

\section{Investigated Setup}
\label{section_setup}
The investigated setup consists of a silicon waveguide with square shape
which is perforated with a number of cylindrical air pores (PhC) and is supported by a silica 
substrate. 
The cavity essentially is formed by a missing central pore. 
Diameters and positions of the air pores are fine-tuned for high 
quality and transmission~\cite{Velha2006a}.
For the FEM simulations we discretize this geometry by prismatoidal elements. 
Figure~\ref{fig_grid} shows a part of the spatial mesh.
A typical computational domain has a size of 
$1.6\,\mu\textrm{m} \times 6\,\mu\textrm{m} \times 1.6\,\mu\textrm{m}$ corresponding 
to roughly $3.5\times 13\times 3.5\approx160$ cubic wavelengths (in Si).
The model extends waveguide, substrate and surrounding air to infinity by  applying 
transparent boundary conditions to the corresponding faces. 
The material relative permittivities are $\epsilon_r=3.48^2$ (Si) and $\epsilon_r=1.44^2$ (SiO$_2$), 
the waveguide width and height are $w=520\,$nm and $h=340\,$nm.
Air holes with diameters 
$d_1 = 130\,$nm, 
$d_2 = 160\,$nm, 
$d_3 = 185\,$nm, 
$d_4 = 181\,$nm, 
$d_N = 181\,$nm (holes with $d_N$ repeat $N=2\dots4$ times), 
are positioned symmetrically around the cavity center with 
center to cavity-center distances $\pm p_i$:
$p_1 = L/2+d_1/2$ (where $L$ is the cavity length, $L=422\dots425\,$nm),
$p_2 = p_1+300\,$nm,
$p_3 = p_2+315\,$nm,
$p_4 = p_3+325\,$nm,
$p_5 = p_4+352\,$nm,
$p_{N+1} = p_{N}+370\,$nm.
This setup has been investigated experimentally and theoretically by 
Velha {\it et al} for $L=425\,$nm~\cite{Velha2006a}.
We have extended the investigation to a fine sampling of 
cavity lengths $L$ on a nanometer scale, and we investigate pore sidewall angles 
different from 90\,degrees. 

For simulating transmission through the microcavity at specific wavelength $\lambda_0$, 
we first compute the fundamental 
propagation mode of the waveguide at $\lambda_0$ using a FEM  
propagating mode solver~\cite{PoZsKlx07}. 
The obtained mode field is applied as input data to one of the boundaries of the 3D 
computational domain (left boundary in Fig.~\ref{fig_grid}), such that the mode propagates 
in direction of the center of the waveguide. 
We then compute the scattered light field in the setup corresponding to this excitation 
using higher-order finite-elements. 
In post-processes we extract energy fluxes through interfaces and field distributions 
in several cross-sections from the 3D light field distribution.
Figure~\ref{fig_solution} visualizes the computed electric field intensity at an incident wavelength 
close to the cavity's resonance wavelength.

\section{Numerical Results}
\begin{figure}[t]
\begin{center}
  \includegraphics[width=.48\textwidth]{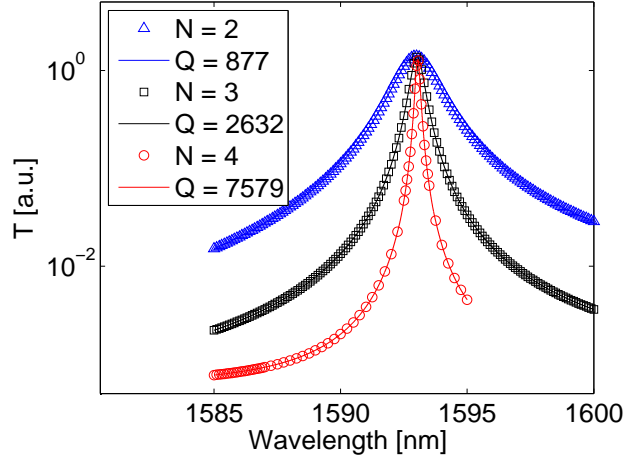}
  \caption{
Transmission spectra for PhC cavities with $L=425\,nm$ and  $N=2/3/4$ on a logarithmic scale. 
}
\label{fig_resonances_velha}
\end{center}
\end{figure}

\subsection{Validation of FEM results}
\label{section_validation}
In order to compare results obtained with our method to results obtained with a 
dedicated Fourier modal method (RCWA)~\cite{Cao2002a,Velha2006a} and to 
experimental results~\cite{Velha2006a} we have performed 
wavelength scans over cavity resonances for cavity lengths 
$L=425\,$nm~\cite{Burger2009Tacona}. 
Figure~\ref{fig_resonances_velha} shows the corresponding transmission spectra for cavities with 
a total number of 12/14/16 PhC holes, resp.~for $N=2/3/4$. 
As expected, for all three cases we find clear resonances in the transmission spectra. 
All resonances are centered around a peak wavelength of about $\lambda_{0,FEM}=1593\,$nm.
From the ratio of  $\lambda_0$ to the widths of Lorentz fits to the distributions we find 
Q-factors of 
$Q_{FEM}=880/2630/7580$ for $N=2/3/4$.
The simulated peak wavelengths correspond very well to experimental results of 
Velha {\it et al} ($\lambda_{0, exp}\approx 1599/1592.7/1592.7\,$nm for $N=2/3/4$)
and to numerical results of 
Velha {\it et al} ($\lambda_{0, RCWA}\approx 1600\,$nm).
The Q-factors obtained with our method are slightly below the experimental and 
numerical values from Velha {\it et al} 
($Q_{exp}=1100/2700/8900$, 
$Q_{RCWA}=1270/3200/9600$). 

To further validate our results we have performed a convergence analysis showing that 
with moderate computational effort in the range of minutes of computation time for a single 
frequency we reach accuracies of the resonance wavelength of better than 0.1\%, and of 
high Q-factors~\cite{Zain2008a} in the few percent range. 

\subsection{Investigation of geometry variations on a nanometer-scale}
\begin{figure}[b]
 \begin{center}
  \includegraphics[width=.56\textwidth]{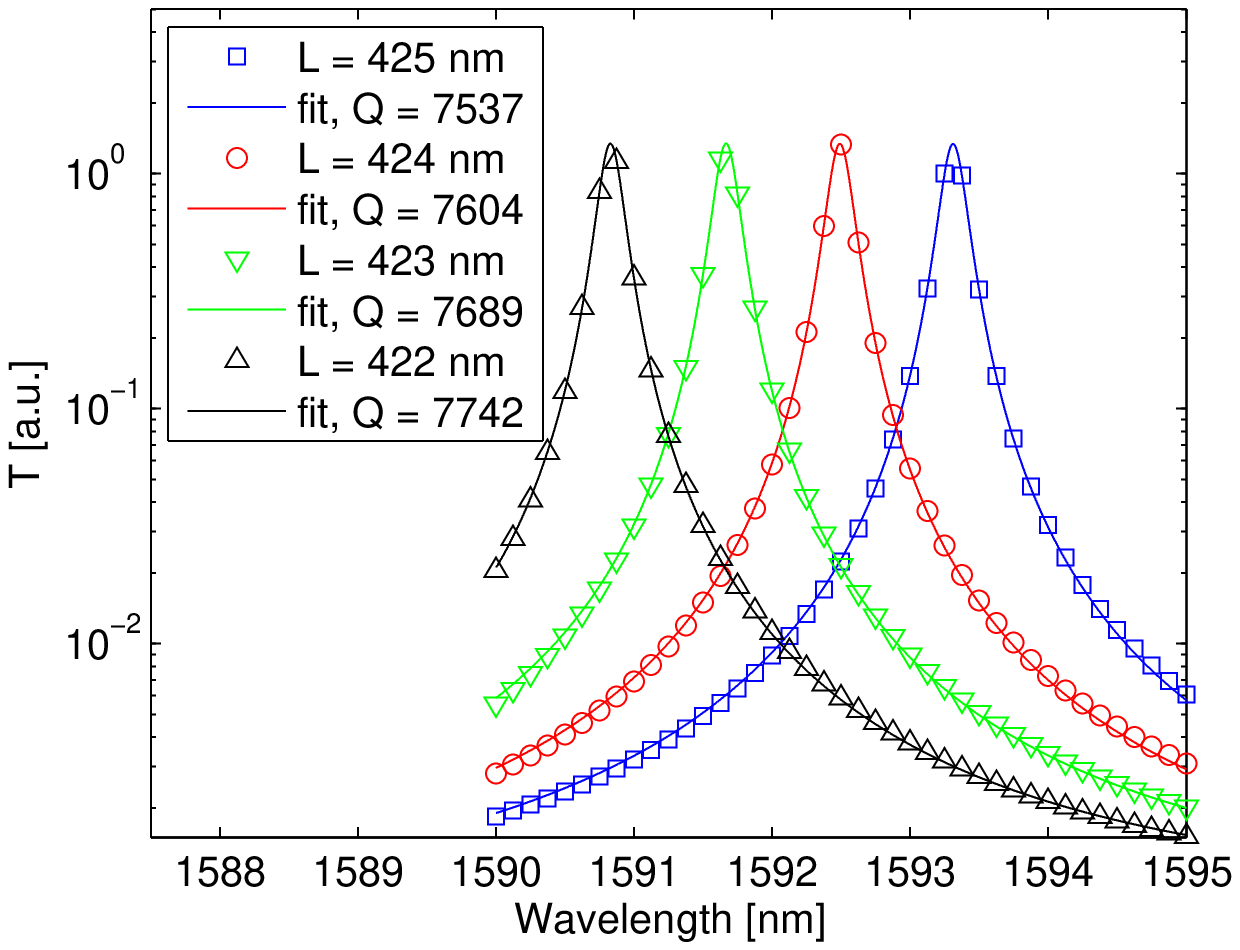}
  \includegraphics[width=.4\textwidth]{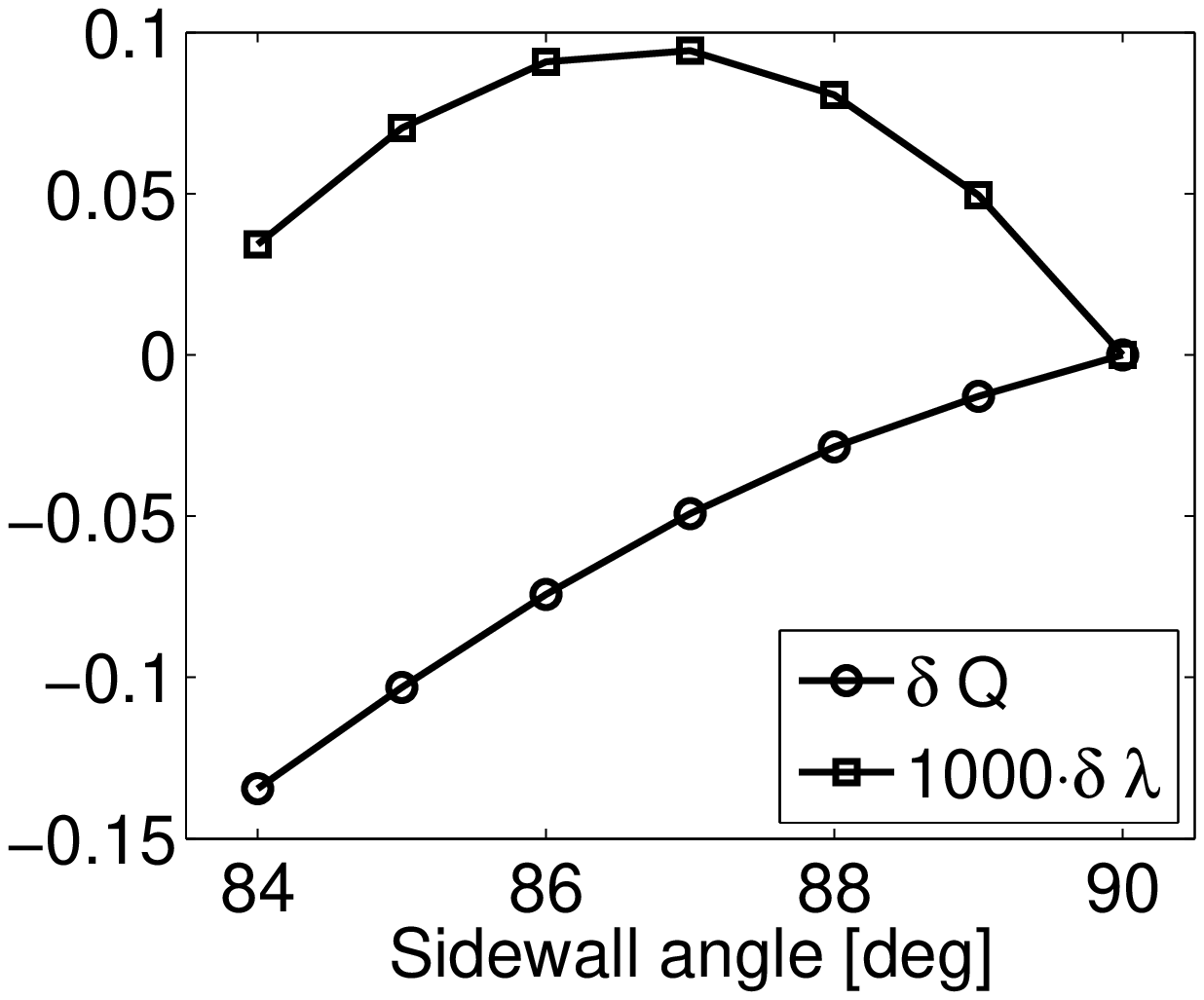}
  \caption{
Left: Transmission spectra for PhC cavities with $L=422\dots 425\,$nm and $N=2$. 
Right: Dependence of relative changes of Q-factor and resonance wavelength with 
sidewall angle 
(for PhC cavities with $L=425\,nm$ and $N=2$).
Circles correspond to relative change of the Q-factor, 
$\delta Q = (Q_{SQA}-Q_{90})/Q_{90}$.
Squares correspond to relative change of resonance wavelength, 
$\delta \lambda = (\lambda_{0,SWA}-\lambda_{0,90})/\lambda_{0,90}$, 
magnified by a factor of 1000.
$Q_{90}$ and $\lambda_{0,90}$ are the values for SWA=90\,deg. 
}
\label{fig_length_scan_swa}
\end{center}
\end{figure}

We have performed numerical experiments with slight variations 
of some geometrical parameters~\cite{Burger2009Tacona}. 
This allows to quantify the influence of fabrication tolerances on 
microcavity properties.
Due to spatial adaptivity of the used unstructured meshes, FEM allows to investigate 
influences of very small variations of the geometry at high accuracy and 
without additional computational costs. 

In a first set of simulations we have performed wavelength scans for PhC cavities with 
12 air pores ($N=2$) for cavity lengths of $L=422, 423, 424, 425\,$nm. 
Figure~\ref{fig_length_scan_swa} (left) shows the corresponding transmission spectra. 
As can be seen from these spectra and from the fitted distributions, 
in the investigated regime, a shift of $L$ shifts the resonance wavelength 
$\lambda_0$ approximately linearly with $\delta \lambda_0/\delta L \approx 0.8$.
Also, a decrease of $L$ by about one percent increases $Q$ by about 4\% in the 
investigated regime. 
This shows that fabrication tolerances in the range of few nanometers 
can well explain significant deviations of device performance. 
The slight shift of resonance wavelength $\lambda_0$ with respect to the 
data displayed in Fig.~\ref{fig_resonances_velha} of about 0.02\% is due 
the lower numerical resolution of the simulations displayed 
in Fig.~\ref{fig_length_scan_swa} (left).

In a second set of experiments we have varied the sidewall angle (SWA) of all air pores 
between 84 and 90 degrees (90~deg.~corresponds to perpendicular SWA, 
resp.~cylindrical shape). For a better comparability of the results, for SWA $<$ 90\,degrees 
the upper radius of 
the air pore was increased, the lower radius was decreased, and the radius at central 
waveguide height was kept constant. 
Figure~\ref{fig_length_scan_swa} (right) shows corresponding results:
While the influence of pore SWA on resonance wavelength can be neglected in the investigated 
regime, the influence of pore SWA on Q-factor is significant: 
A deviation of 5\,degrees from vertical sidewalls decreases the Q-factor 
by about 10\,percent.

\begin{figure}
\begin{center}
  \includegraphics[width=.48\textwidth]{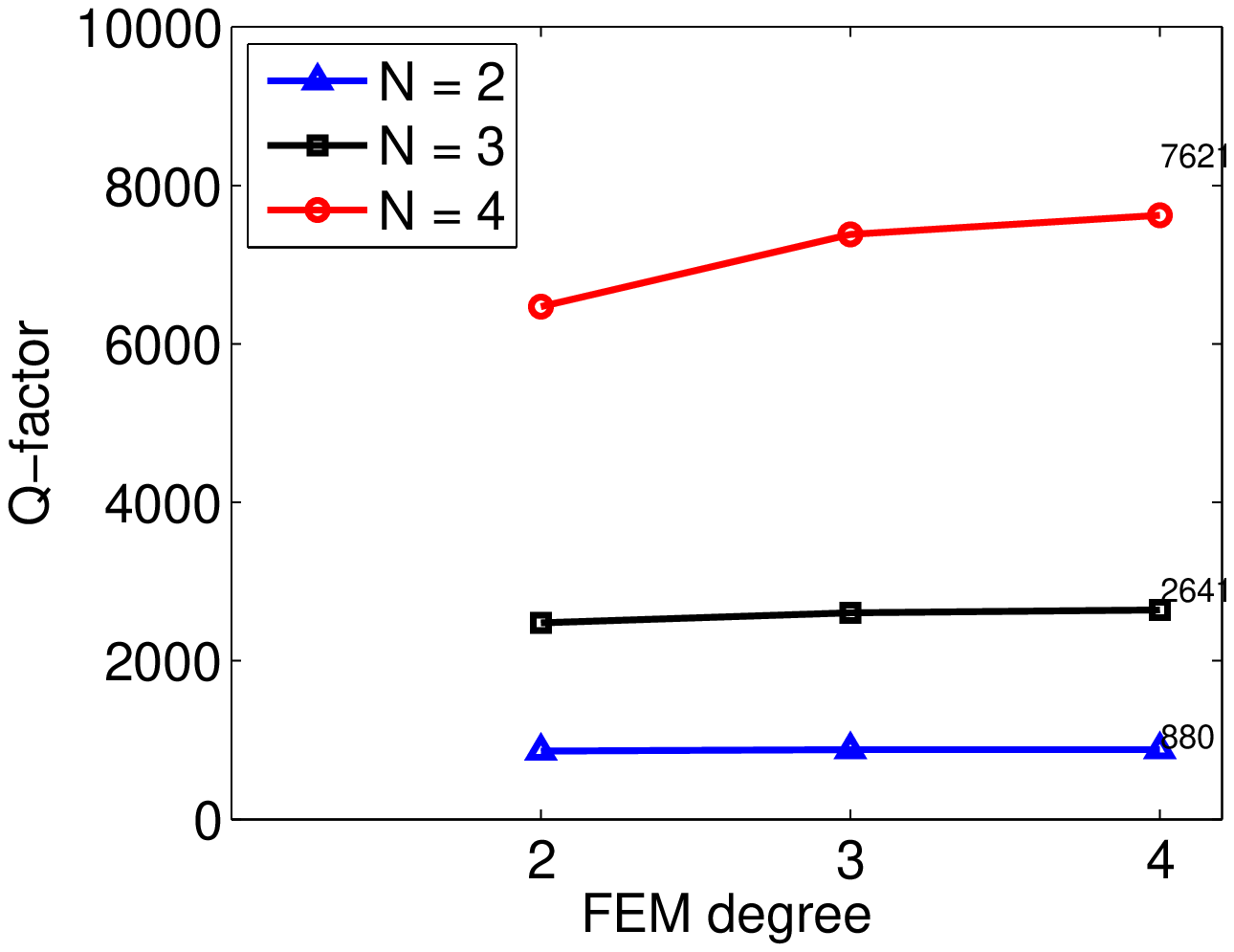}
  \includegraphics[width=.48\textwidth]{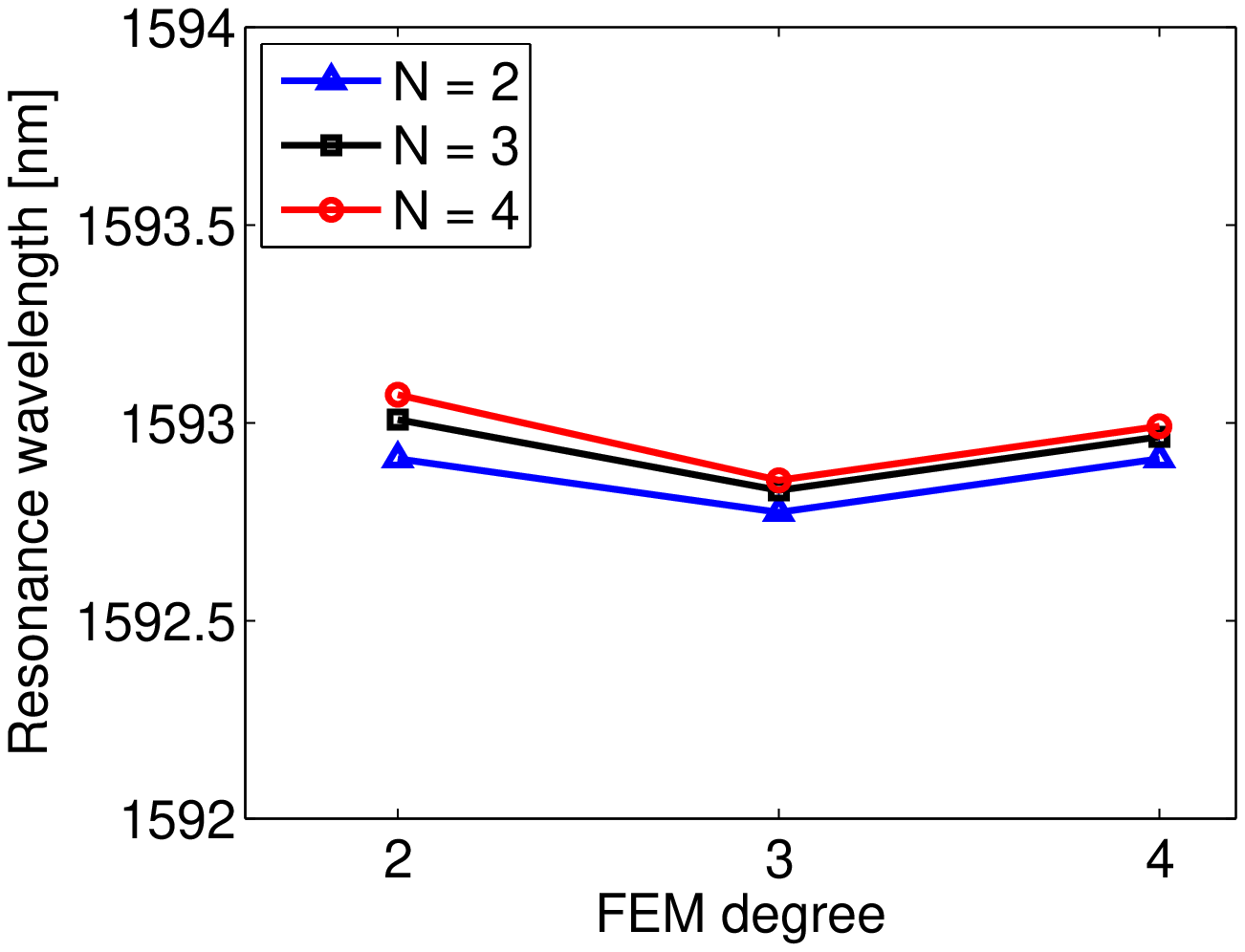}
  \caption{
Convergence of the $Q$-factor (left) and of the resonance wavelength (right) obtained with a 
FEM eigenmode solver for polynomial finite element degree $p=2,3,4$.
}
\label{fig_rs_conv}
\end{center}
\end{figure}

\begin{figure}
\begin{center}
  \includegraphics[width=.48\textwidth]{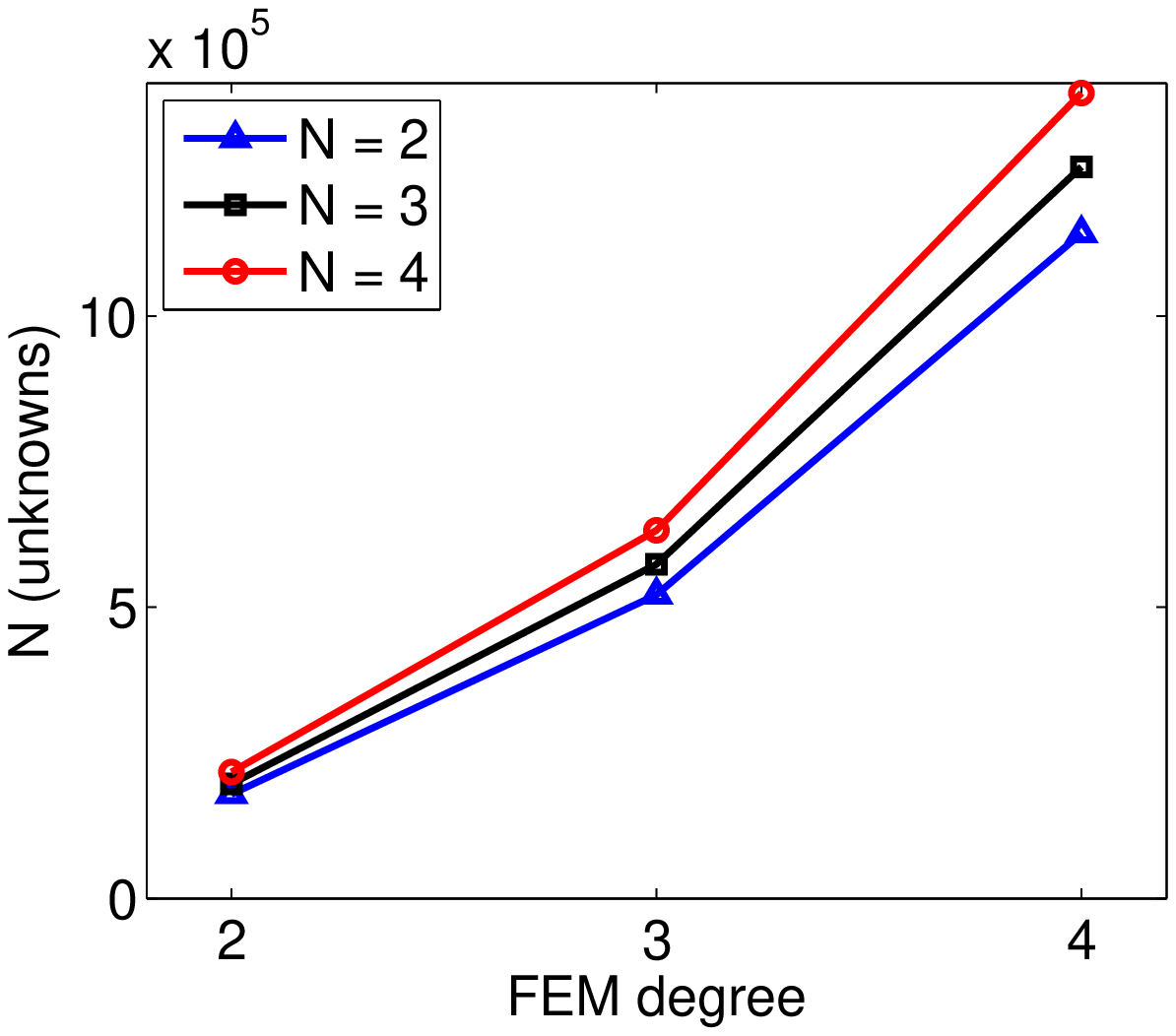}
  \includegraphics[width=.48\textwidth]{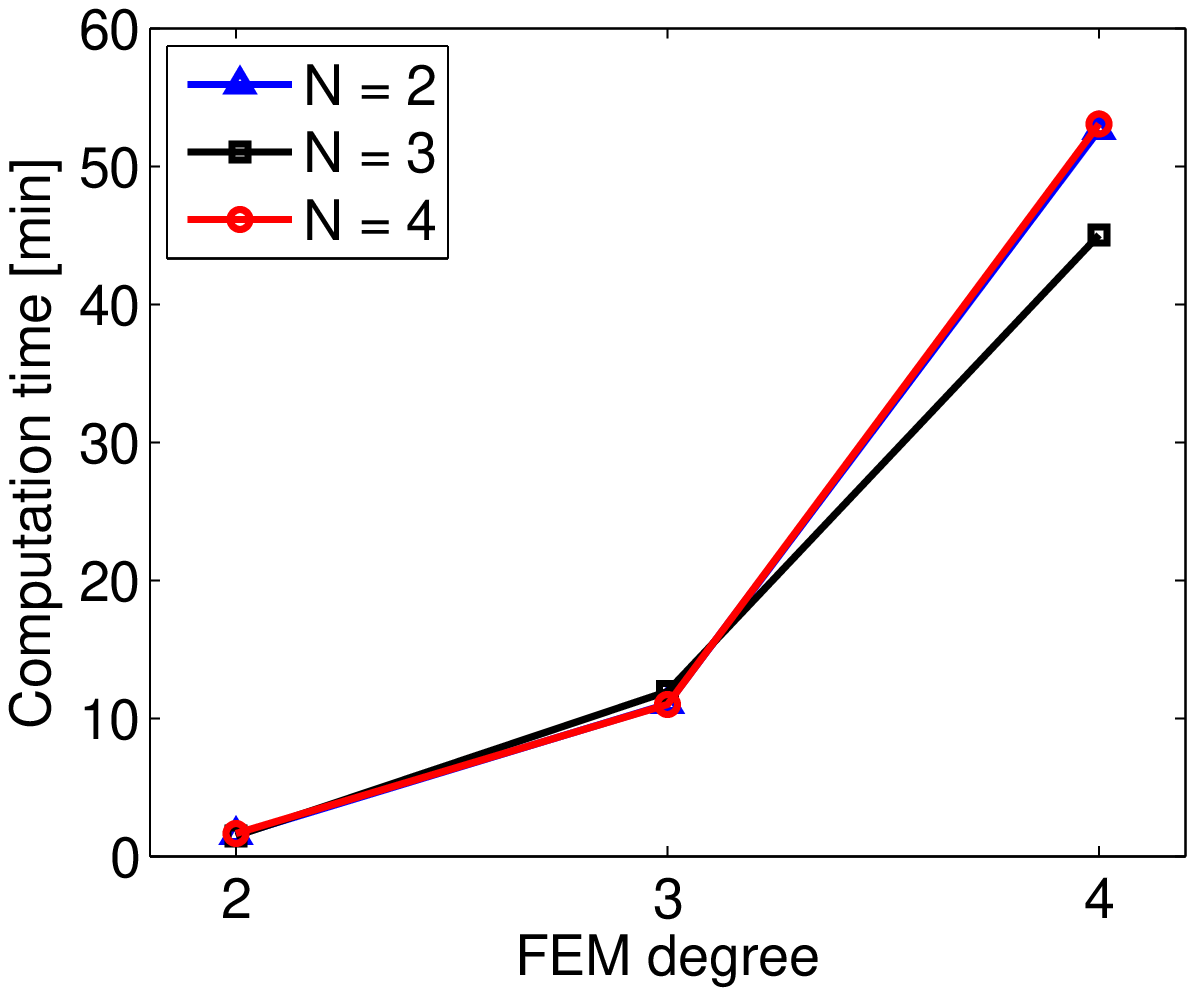}
  \caption{
Computational costs corresponding to the simulation results displayed in 
Figure~\ref{fig_rs_conv}. Left: Number of unknowns in the corresponding system of linear 
equations, $N$. Right: Computation time on a workstation with four double-core 
AMD Opteron CPU's at 2.8\,GHz has been used. Maximum memory consumption was about 
32\,GB of RAM (for N=4, $p=4$).
}
\label{fig_rs_cpu}
\end{center}
\end{figure}

\subsection{Direct simulation of cavity resonances}
Alternatively to simulating the light scattering response of the device, we can also 
use an eigenmode solver for directly simulating the resonance properties (resonance wavelength 
and Q-factor): 
Given the geometrical setup as described in Section~\ref{section_setup}, 
one computes an electric field distribution $E$
and a complex eigenfrequency $\omega$ which satisfy Maxwell's time-harmonic wave equation 
$$\nabla \times \mu^{-1} \nabla \times E = \omega^2 \varepsilon E$$
without sources; 
electric permittivity and magnetic permeability are denoted by $\varepsilon$ and $\mu$, respectively.
Transparent boundary conditions take into account the specific geometry 
of the problem where waveguides are modelled to  
extend to infinity in the exterior domain~\cite{Zschiedrich2006pml}. 
When the eigenmode ($E, \omega$) is computed, the respective $Q$-factor is deduced from the real and imaginary parts 
of the complex eigenfrequency,
$Q=\Re(\omega)/(-2\Im(\omega)),$
the resonance wavelength $\lambda_{0}$ is given by $\lambda_{0}=2\pi c_0/\omega$, with the speed of light $c_0$.

Figure~\ref{fig_rs_conv} shows the convergence of 
the computed $Q$-factors and resonance wavelengths for the cases investigated also 
in Section~\ref{section_validation} ($N=2/3/4$). 
Here we have increased the numerical resolution by increasing the polynomial degree $p$ of the 
chosen finite element ansatz-functions. 
As can be seen from this figure, both, Q-factor and resonance wavelength computed with the 
eigenmode solver converge to the values obtained from the light scattering simulations 
({\it c.f.} Section~\ref{section_validation}). 
Figure~\ref{fig_rs_cpu} shows the computational costs corresponding to these simulations.
Please note that simulating a resonance directly requires a single computation only while 
deducing the resonance properties from a transmission or 
reflection simulation (as in Section~\ref{section_validation})
always requires several computations at various wavelengths. 
Especially for high-Q resonances, where the choice of such wavelengths for a transmission scan is not 
clear a-priori, direct computation of resonances greatly simplifies the simulation task and reduces 
computational effort.

\subsection{Simulation of high-Q resonators}

\begin{figure}
\begin{center}
  \includegraphics[width=.48\textwidth]{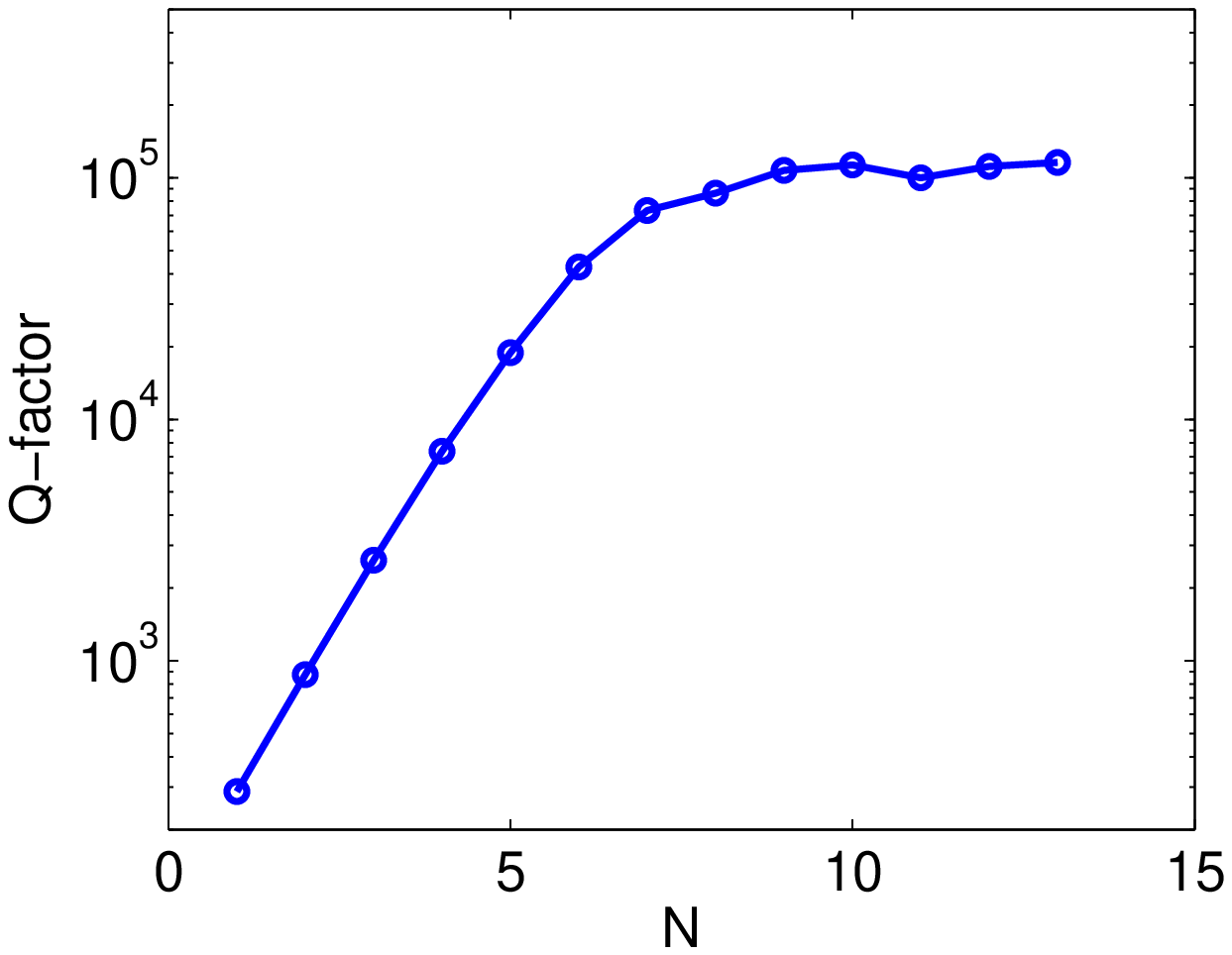}
  \includegraphics[width=.48\textwidth]{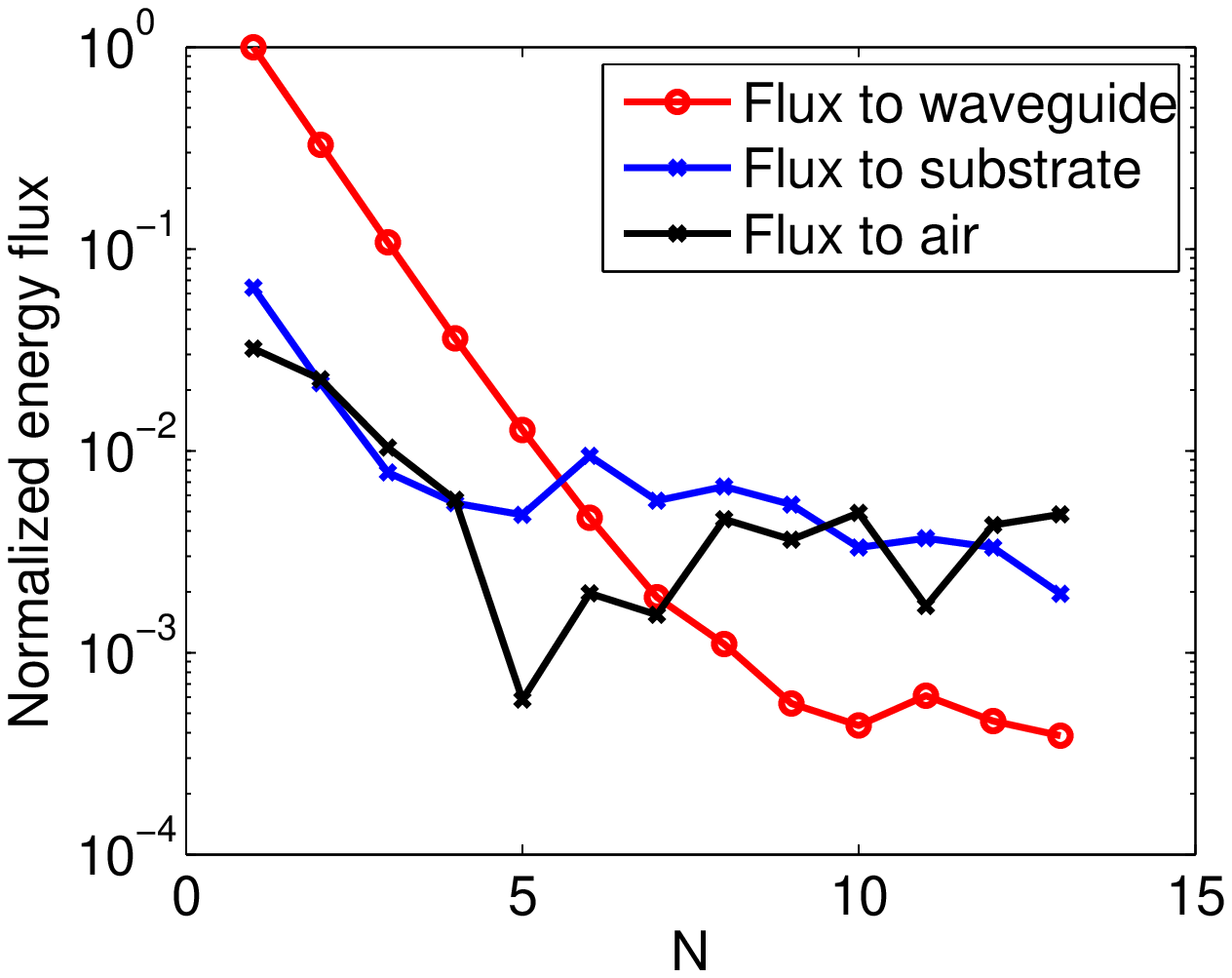}
  \caption{
Left: Q-factor of PhC microcavities in dependence on number of air pores in the periodic sections of 
the design, $N$. 
Right: Normalized energy flux to waveguide, substrate, air regions 
in dependence on number of air pores in the periodic sections of 
the design, $N$. 
}
\label{fig_rs_nr}
\end{center}
\end{figure}

In a 3D photonic crystal setup one expects that the cavity Q-factor will increase exponentially 
with the width of the surrounding photonic crystal when a cavity is operated inside a bandgap. 
However, in the given 1D photonic crystal setup, the situation is different because leakage to 
the substrate and to surrounding air regions also have to be taken into account. 
In a further set of numerical experiments we have applied the eigenmode FEM  solver in order to 
adress the question which Q-factors can be reached by simply adding more and more pores to the 
1D PhC microcavity. 
Figure~\ref{fig_rs_nr} (left) shows that the cavity Q-factor increases exponentially with 
the size of the photonic crystal region for numbers of pores  in the periodic sections of the 
design, $N$, between 1 and about 6. For higher $N$, a plateau of $Q \sim 100,000$ is reached. 

By integrating over the flux of energy density over the 
computational domain boundaries in the different 
material regions we have further recorded the losses to the substrate, to the surrounding air region, 
and to the waveguide region. 
Figure~\ref{fig_rs_nr} (right) shows the corresponding results. 
In accordance with Fig.~\ref{fig_rs_nr} (left), for $N$ below $\sim 6$, the losses to substrate and 
surrounding air region do not play a significant role (note the logarithmic scale of the plot). 
As in this regime the resonator looses its energy mainly through the waveguide, the photonic crystal 
perforation of the waveguide yields an exponential decay of the field and exponentially lower losses 
with increasing the width of the waveguide. 
For $N>6$ light scattering to the substrate and to the surrounding air regions are the dominant loss 
channels, and further improvement of the photonic crystal in the waveguide region does therefore 
not lead to an improvement of the quality factor beyond the values reached at $N\sim 6$.

\section{Conclusion}
Finite-element solvers have been used to numerically investigate photonic crystal microcavities. 
The results have been validated by comparison to experimental and theoretical results from 
the literature. 
Further, numerical solutions from a scattering solver and numerical solutions from an eigenmode solver have been 
shown to yield an excellent agreement of physical results within the limits of numerical errors. 
The influence of nanometer-scale geometry variations (placement and sidewall angle) 
on the performance of one-dimensional PhC microcavities has been investigated. 
Losses to substrate and air regions have been identified as limiting factors to 
ultra-high Q-factor performance of the investigated specific photonic crystal microcavity design.

\section*{Acknowledgments}
We acknowledge support by the German Federal Ministry of
Education and Research, BMBF, under contract No.\,13N9071. 

\bibliography{/home/numerik/bzfburge/texte/biblios/phcbibli,/home/numerik/bzfburge/texte/biblios/group,/home/numerik/bzfburge/texte/biblios/lithography}

\begin{thebibliography}{10}

\bibitem{Zain2008a}
Zain, A. R.~M., Johnson, N.~P., Sorel, M., and {De La Rue}, R.~M., ``Ultra high
  quality factor one dimensional photonic crystal/photonic wire micro-cavities
  in silicon-on-insulator ({SOI}),'' {\em Opt. Express}~{\bf 16}(16),  12084
  (2008).

\bibitem{Tanaka2008a}
Tanaka, Y., Asano, T., and Noda, S., ``Design of photonic crystal nanocavity
  with {Q}-factor of $\sim 10^9$,'' {\em J. Lightwave Technol.}~{\bf 26}(11),
  1532--1539 (2008).

\bibitem{Ripin2000a}
Ripin, D.~J., Lim, K.-Y., Petrich, G.~S., Villeneuve, P.~R., Fan, S., Thoen,
  E.~R., Joannopoulos, J.~D., Ippen, E.~P., and Kolodziejski, L.~A., ``Photonic
  band gap airbridge microcavity resonances in {GaAs/Al$_x$O$_y$} waveguides,''
  {\em Journal of Applied Physics}~{\bf 87}(3),  1578--1580 (2000).

\bibitem{Schoenenberger2009a}
Sch\"{o}nenberger, S., Moll, N., St\"{o}ferle, T., Mahrt, R.~F., Offrein,
  B.~J., G\"{o}tzinger, S., Sandoghdar, V., Bolten, J., Wahlbrink, T.,
  Pl\"{o}tzing, T., Waldow, M., and F\"{o}rst, M., ``Circular grating
  resonators as small mode-volume microcavities for switching,'' {\em Opt.
  Express}~{\bf 17}(8),  5953 (2009).

\bibitem{Burger2009Tacona}
Burger, S. and Zschiedrich, L., ``Numerical investigation of
  silicon-on-insulator {1D} photonic crystal microcavities,'' in [{\em
  Theoretical and computational nanooptics: Proceedings of the 2nd
  International Workshop}{\nolinebreak\hspace{0.1em}]},  Chigrin, D.~N., ed.,
  {\bf 1176},  43--45, AIP (2009).

\bibitem{Burger2008ipnra}
Burger, S., Zschiedrich, L., Pomplun, J., and Schmidt, F., ``{JCMsuite}: {A}n
  adaptive {FEM} solver for precise simulations in nano-optics,'' in [{\em
  Integrated Photonics and Nanophotonics Research and
  Applications}{\nolinebreak\hspace{0.1em}]},   ITuE4, Optical Society of
  America (2008).

\bibitem{Velha2006a}
Velha, P., Rodier, J.~C., Lalanne, P., Hugonin, J.~P., Peyrade, D., Picard, E.,
  Charvolin, T., and Hadji, E., ``Ultra-high-reflectivity photonic-bandgap
  mirrors in a ridge {SOI} waveguide,'' {\em New Journal of Physics}~{\bf
  8}(9),  204 (2006).

\bibitem{PoZsKlx07}
Pomplun, J., Zschiedrich, L., Klose, R., Schmidt, F., and Burger, S., ``Finite
  element simulation of radiation losses in photonic crystal fibers,'' {\em
  phys. stat. sol. (a)}~{\bf 204},  3822 (2007).

\bibitem{Cao2002a}
Cao, Q., Lalanne, P., and Hugonin, J.-P., ``Stable and efficient {B}loch-mode
  computational method for one-dimensional grating waveguides,'' {\em J. Opt.
  Soc. Am. A}~{\bf 19}(2),  335 (2002).

\bibitem{Zschiedrich2006pml}
Zschiedrich, L., Klose, R., Sch\"adle, A., and Schmidt, F., ``A new finite
  element realization of the {P}erfectly {M}atched {L}ayer {M}ethod for
  {H}elmholtz scattering problems on polygonal domains in 2{D},'' {\em J.
  Comput. Appl. Math.}~{\bf 188},  12--32 (2006).

\end{thebibliography}
\bibliographystyle{spiebib}  

\end{document}